\documentclass[aps,prl,reprint,superscriptaddress]{revtex4-2}

\usepackage{amsmath}
\usepackage{amsthm}
\usepackage{amssymb}
\usepackage{amsfonts}
\usepackage{blindtext}
\usepackage{bm}
\usepackage{amsfonts}
\usepackage{graphicx}
\usepackage{amssymb}
\usepackage{breqn}
\usepackage{stackengine}
\usepackage{diagbox}
\usepackage{booktabs}
\usepackage[normalem]{ulem}
\usepackage{cancel}
\usepackage{circledtext}
\usepackage{subfigure}

\newcommand{\ket}[1]{| #1 \rangle}

\newcommand{\avg}[1]{\left\langle #1 \right\rangle}

\newcommand{\kvec}{\mathbf{k}}

\newcommand{\Tr}{\mathrm{Tr}}

\makeatletter
\let\cat@comma@active\@empty
\makeatother

\begin{document}


\title{Many-body Josephson diode effect in superconducting quantum interferometers}



\author{Zelei Zhang}
\affiliation{Shanghai Institute of Microsystem and Information Technology,
Chinese Academy of Sciences, Shanghai 200050, China}
\affiliation{School of Physical Science and Technology, ShanghaiTech University, Shanghai 201210, China}

\author{Jianxiong Zhai}
\affiliation{Shanghai Institute of Microsystem and Information Technology,
Chinese Academy of Sciences, Shanghai 200050, China}
\affiliation{University of Chinese Academy of Sciences, Beijing 100049, China}

\author{Yi Zhang}
\email{zhangyi821@shu.edu.cn}
\affiliation{Department of Physics and Institute for Quantum Science and Technology, Shanghai University, Shanghai 200444, China}
\affiliation{Shanghai Key Laboratory of High Temperature Superconductors, Shanghai University, Shanghai 200444, China}

\author{Jiawei Yan}
\email{yanjw@mail.sim.ac.cn}
\affiliation{Shanghai Institute of Microsystem and Information Technology,
Chinese Academy of Sciences, Shanghai 200050, China}
\affiliation{University of Chinese Academy of Sciences, Beijing 100049, China}


\date{\today}

\begin{abstract}
We propose a many-body mechanism for a strong Josephson diode effect (JDE) in an interacting nanoscale SQUID formed by two parallel quantum dots coupled to superconducting leads.
Unlike conventional diode behavior, where nonreciprocity originates from a skewed current-phase relation within a single, continuously evolving ground state, the JDE reported here is \emph{branch selected}: the positive and negative critical currents are optimized on different many-body branches across the $0$-$\pi$ phase boundary, yielding a substantial enhancement of the diode efficiency.
We further show that a \emph{nonlocal} Cooper-pair tunneling channel, which binds the two electrons on different arms, is essential: it reshapes the $0$-$\pi$ boundary and produces a pronounced ``diode band'' in parameter space, in sharp contrast to the fragile hotspot obtained when only local Cooper-pair transfer is available.
While the key physics is captured by an effective model in the superconducting atomic limit, our conclusions remain robust for realistic finite-gap devices, as demonstrated within a generalized atomic-limit framework.
\end{abstract}


\hyphenation{single}

\maketitle

\paragraph{Introduction --}
Nonreciprocal superconducting transport, manifested by direction-dependent critical currents ($I_{c+}\neq I_{c-}$), provides a route to dissipationless rectification and has emerged as a central topic in superconducting electronics and hybrid quantum devices \cite{Kun2022,Nadeem2023,Daido2022,Davydova2022,Zhang2022}.
Following its initial observation in noncentrosymmetric superconductors \cite{Ando2020,Narita2022}, diode-like supercurrents have been reported in a wide range of systems, including few-layer superconductors \cite{Bauriedl2022,DiezMerida2023}, van der Waals heterostructures \cite{Idzuchi2021,Wu2022,Kim2024}, and Rashba/topological and quantum-dot Josephson junctions \cite{Pal2022,Turini2022,Trahms2023}.
Among them, Josephson junctions are particularly attractive because their current-phase relation (CPR) can be highly engineered \cite{Souto2022,Senapati2023,Reinhardt2024,Li2024}, making them a versatile platform for superconducting nonreciprocity.

\begin{figure}
\includegraphics[width=1.0\linewidth]{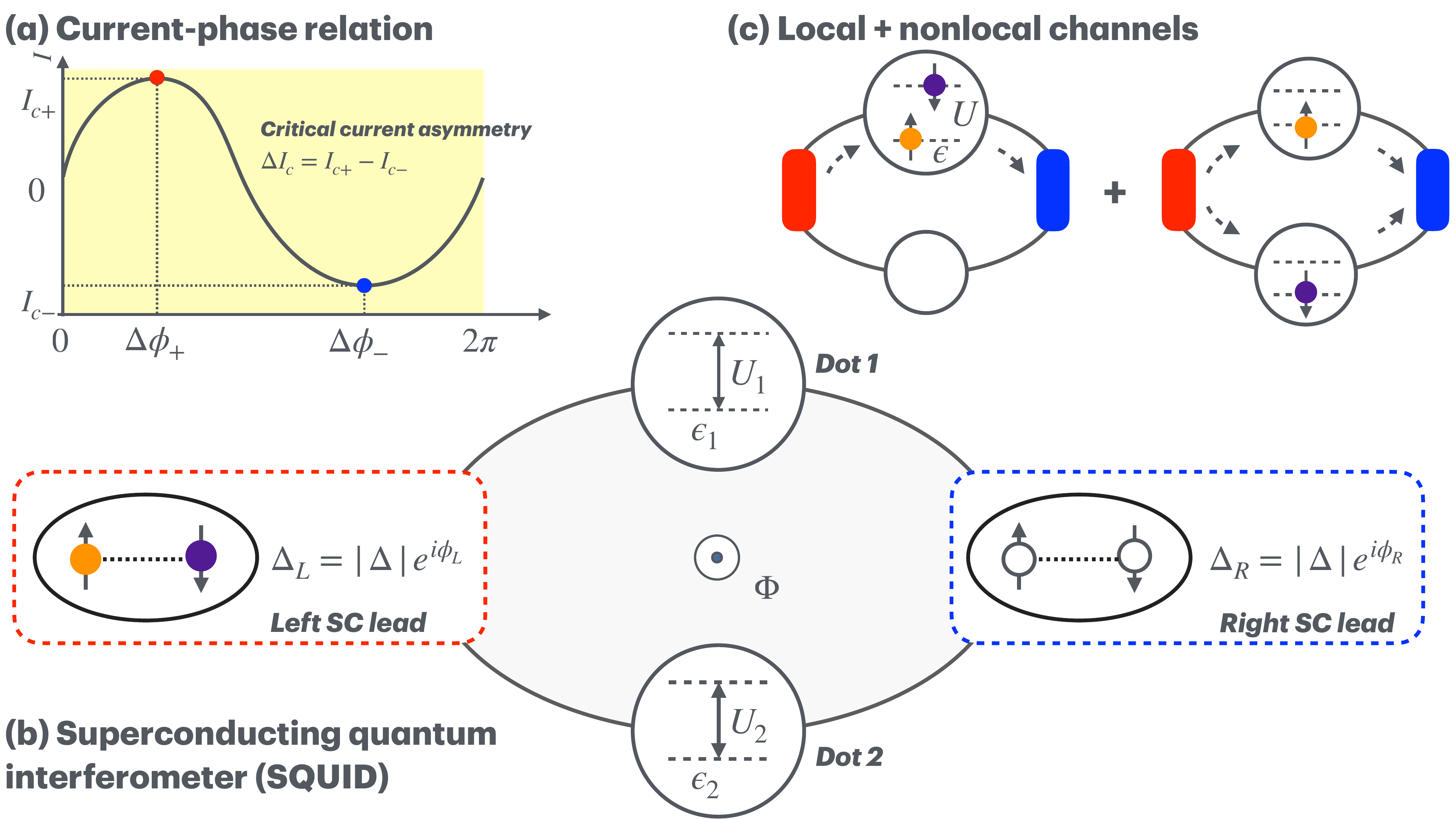}
\caption{(a) Current phase relation (CPR) and critical currents $I_{c\pm} = \max \{ \pm I(0 \le \Delta \phi < 2\pi) \}$. (b) Schematic of the SQUID model, which contains two quantum dots connected in parallel to the superconducting leads.  (c) Local and non-local transport channels for Cooper pairs tunneling.}
\label{fig: squids model and spectra}
\end{figure}

Theoretically, generating the Josephson diode effect (JDE) generally requires the simultaneous breaking of both inversion symmetry (IS) and time-reversal symmetry (TRS) \cite{Buzdin2008,He2022phenomenological,Zhang2022,Wu2025}.
Established mechanisms include finite-momentum pairing \cite{Yuan2022,Davydova2022,Pal2022}, spin-orbit coupling combined with Zeeman or exchange fields \cite{Ando2020,Iliifmmodeacutecelsecfi2022,Bau_2022,Costa2023,Debnath2024}, multichannel interferometry \cite{Souto2022,Fominov2022,Lu2023,Gupta2023} and loop-current states~\cite{varma2025,zhang2025}.
Recent studies further suggest that diode responses can be amplified by chiral band structures \cite{He2023,Cheng2023}, Majorana modes \cite{Tanaka2022,Legg2023,cayao2024enhancing}, and competing ordered states \cite{Banerjee2024}. 
However, most existing studies are confined to the weakly interacting regime, where the ground state (GS) evolves smoothly within a single many-body branch and rectification mainly originates from a skewed CPR.
By contrast, strong onsite Coulomb repulsion can drive a parity-changing quantum phase transition between competing many-body states, known as the $0$-$\pi$ transition \cite{Rozhkov1999,Siano2004,Dam2006,Maurand2012,Bargerbos2022,Meden2019}.
At zero temperature, such a transition can induce non-analytic changes in the free energy $F = -k_B T \ln Z$, so that the Josephson current
\begin{equation}\label{eq: Josephson current}
I(\Delta\phi)=\frac{2e}{\hbar}\left\langle \frac{\partial \hat{H}}{\partial \Delta\phi}\right\rangle
=\frac{2e}{\hbar}\frac{\partial F}{\partial \Delta\phi}~,
\end{equation}
with $\Delta \phi := \phi_L - \phi_R$ the phase difference between the left and right leads, can change abruptly across the transition point.
This suggests a qualitatively different route to superconducting nonreciprocity: rather than merely distorting the CPR within a single branch, interactions may allow the positive- and negative-direction critical currents, defined by $I_{c\pm}=\max\{\pm I(0\le \Delta\phi<2\pi)\}$ see Fig.~\ref{fig: squids model and spectra}(a), to be selected from different many-body branches near a $0$-$\pi$ boundary.

A minimal setting for exploring this interaction-driven JDEs is an asymmetric superconducting quantum interference device (SQUID) comprising two parallel quantum dots coupled to two superconducting leads, as depicted in Fig.~\ref{fig: squids model and spectra}(b).
In this geometry, TRS is broken by a magnetic flux threading the SQUID loop, while IS is lifted by making the two dots inequivalent, for instance through dot detuning \cite{Debbarma2023,Ciaccia2023} or by coupling to a non-Hermitian environment \cite{Qi2025}.
Moreover, nanoscale SQUIDs naturally support transport channels beyond local Cooper pair transfer through an individual dot.
In particular, a nonlocal pair-splitting channel allows the two electrons of a Cooper pair to traverse different dots, forming a nonlocal pairing bound state, see Fig.~\ref{fig: squids model and spectra}(c)\cite{Recher2001,Hofstetter2009,Wang2011,Deacon2015,Herrmann2010,Fueloep2015}.
This process is expected to become prominent at strong $U$, where local pair tunneling is strongly suppressed, and therefore must be treated explicitly when analyzing interacting SQUIDs.

In this Letter, we demonstrate that an interacting double-dot SQUID can host a many-body \emph{branch selected} JDE in which the two critical currents are controlled by different many-body branches near the $0$-$\pi$ boundary.
Using an effective Hamiltonian in the superconducting atomic limit, we establish two central results:
(i) an interaction-enabled diode mechanism where $I_{c+}$ and $I_{c-}$ are selected from distinct ($0$ and $\pi$) branches at their respective phase extrema, producing strongly enhanced rectification;
(ii) the \emph{nonlocal} Cooper pairing channel acts as a control knob by reducing the energy splitting between competing nonlocal $0$ and $\pi$ states, reshaping the phase boundary, and converting a narrow, parameter-sensitive diode region (``diode hotspot'') into a robust ``diode band''.
The applicability of our results to a more general finite-gap model also persists, as we further discuss within a generalized atomic-limit framework.
Hereafter, we set $e = \hbar = k_B = 1$.

\paragraph{Minimal Model --}

The minimal Hamiltonian capturing the essential SQUID physics reads (see Fig.~\ref{fig: squids model and spectra} (a))
\begin{equation}\label{eq: minimal model Hamiltonian}
\hat{H}^{\rm eff} =\sum_{i=1}^{2}\big(\hat{H}_i^{\mathrm{dot}}+\hat{H}_i^{\mathrm{loc}}\big)+ \zeta \hat{H}^{\mathrm{nl}}~,
\end{equation}
where
$\hat{H}_i^{\mathrm{dot}}
=\sum_{\sigma}\epsilon_{i\sigma} d^\dagger_{i\sigma} d_{i\sigma}
+ U_i \hat n_{i\uparrow}\hat n_{i\downarrow}$ with $d_{i\sigma}^{(\dagger)}$ annihilates (creates) an electron with spin $\sigma$ on dot $i$, and $\hat n_{i\sigma}=d_{i\sigma}^\dagger d_{i\sigma}$.
$\epsilon_{i\sigma}$ denotes the spin-dependent onsite energy and $U_i$ the on-site Coulomb charging energy of dot $i$.
Superconducting proximity effects from the leads enter in two distinct ways:
(i) $\hat{H}_i^{\mathrm{loc}} =\alpha_i (d_{i\uparrow}^\dagger d_{i\downarrow}^\dagger+\mathrm{H.c.} )$ accounts for \emph{local} pairing, where both electrons of a Cooper pair occupy the same dot.
Its amplitude $\alpha_{1,2}(\Phi,\Delta\phi)=-2\Gamma\cos \left( (\Delta\phi \pm \Phi)/2 \right)$ depends on the superconducting phase difference $\Delta\phi := \phi_L - \phi_R$ and on the magnetic flux $\Phi$ threading the SQUID loop, with $\Gamma = \pi | t^0 |^2 \bar{\rho}$ the effective dot-lead coupling strength ($t^0$ for tunneling between the lead and the dot without magnetic field, and $\bar{\rho}$ for averaged density of states in the leads).
(ii) $\hat{H}^{\mathrm{nl}}
= \beta (d_{1\uparrow}^\dagger d_{2\downarrow}^\dagger
+d_{2\uparrow}^\dagger d_{1\downarrow}^\dagger+\mathrm{H.c.} )$
describes \emph{nonlocal} pairing, where a Cooper pair is split between the two dots, forming an inter-dot bound state.
The corresponding amplitude $\beta(\Delta\phi)=-2\Gamma\cos \left(\Delta\phi / 2\right)$ depends only on $\Delta\phi$, 
since the two electrons of a split Cooper pair traverse opposite arms of the loop, their Aharonov-Bohm phases cancel upon summation \cite{Wang2011}.
Hereafter, we set $\Gamma = 1$ as the unit of energy.
$\zeta\in[0,1]$ quantifies the relative weight of nonlocal pairing compared to local proximity pairing, which depends on the separation between the two arms relative to the superconducting coherence length in the leads \cite{Wang2011}.
Accordingly, nonlocal pairing becomes particularly important in compact devices with small arm separation.
In the Supplemental Material (SM), we derive the Hamiltonian in Eq.~\eqref{eq: minimal model Hamiltonian} starting from a microscopic model, and then consider the limit of an infinite superconducting gap amplitude, $|\Delta| \to \infty$.

Since the Hamiltonian in Eq.\eqref{eq: minimal model Hamiltonian} contains only the density and the singlet pairing terms, which implies \emph{parity} $\hat{\Pi} = (-1)^{\hat{N}}$ and \emph{spin-z projection} $\hat{S}_z$ symmetries.
As results, the full Hamiltonian in Eq.\eqref{eq: minimal model Hamiltonian} takes a block diagonal form (see SM for details)
\begin{dmath}\label{eq: Hamiltonian sector}
\hat{H}_{S_z = 0}^{\Pi=1} \oplus \hat{H}_{S_z =1}^{\Pi =1} \oplus \hat{H}_{S_z=-1}^{\Pi =1} \oplus \hat{H}_{S_z=1/2}^{\Pi=-1} \oplus \hat{H}_{S_z = -1/2}^{\Pi=-1}
\end{dmath}
from which one can exactly diagonalize each sector, yielding the many-body states and thus all the physical quantities can be, in principle, calculated.

\paragraph{Josephson diode effect --}

In our SQUID setup, TRS and IS are broken by magnetic flux $\Phi$ threading the ring and dot detuning $d\epsilon := -2\epsilon_1= 2\epsilon_2$, respectively, providing the essential ingredients for JDE.
Before presenting the numerical results, it is physically clear that for large $d\epsilon$ the system reduces to a single-dot setup\cite{MartinRodero2011,Zonda2015} and the diode is expected to vanish due to the recovery of the IS in this limit.

\begin{figure}
\includegraphics[width=1.0\linewidth]{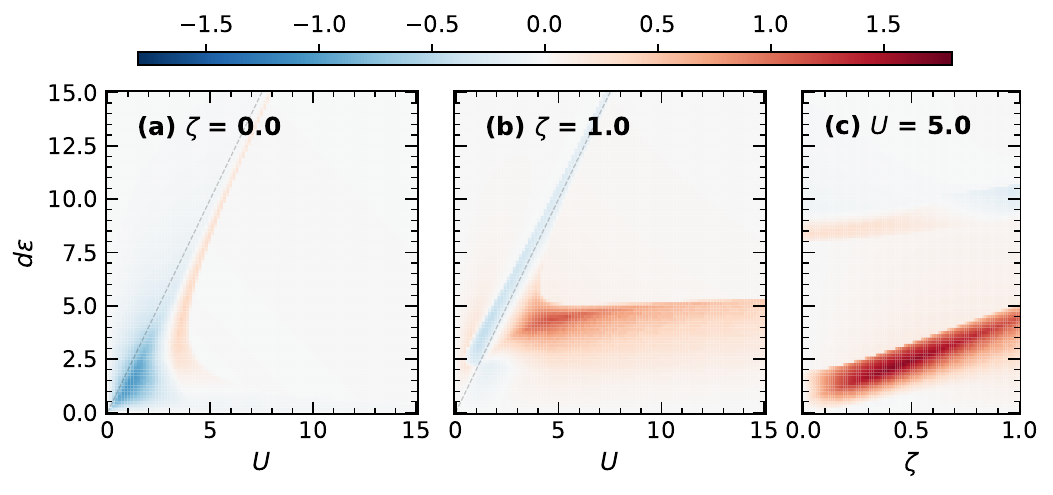}
\caption{Absolute critical current asymmetry $\Delta I_c = I_{c+} - I_{c-}$ in the $(d\epsilon,U)$-plane for (a) $\zeta = 0.0$ and (b) $\zeta = 1.0$, respectively. (c) $\Delta I_c$ in $(d\epsilon,\zeta)$-plane at $U=5.0$. Calculations are performed at zero temperature with $\Phi = 1.13$. Nonlocal Cooper pairing continuously develops a significant ``diode band'' over a wide interaction range, see main text.}
\label{fig: eta phase}
\end{figure}

\begin{figure*}
\includegraphics[width=1.0\linewidth]{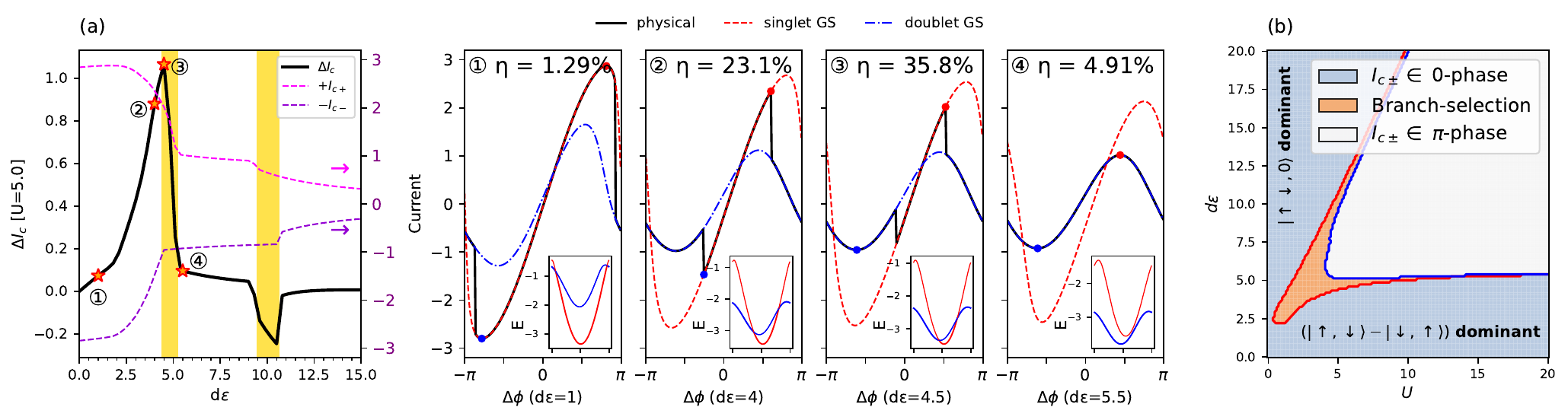}
\caption{(a) $\Delta I_c = I_{c+} - I_{c-}$ (black solid line) vs detuning $d\epsilon$ for $U = 5.0$ at zero temperature ($\zeta = 1.0$ model).
The magenta and violet dashed lines are for $\pm I_{c\pm}$, respectively. The diode peak evolves through four distinct stages, labeled \circledtext{1} to \circledtext{4}, with the corresponding CPRs displayed in the middle panels. Insets depict the GS energy for the singlet and double states. (b) Ground state sectors for $I_{c\pm}$: Lightblue and grey regions indicate where both $I_{c\pm}$ reside in the $0$- and $\pi$-phase, respectively. 
The orange region highlights the branch selected regime, where $I_{c+}$ and $I_{c-}$ are located on different branches.}
\label{fig: mechanism}
\end{figure*}

Fig.~\ref{fig: eta phase}(a) and (b) display the absolute critical current asymmetry $\Delta I_c = I_{c+} - I_{c-}$ in the $(d\epsilon,U)$-plane at fixed magnetic flux $\Phi = 1.13$ and zero temperature, for $\zeta = 0.0$ (without nonlocal pairing) and $1.0$ (with nonlocal pairing), respectively.
We focus on $\Delta I_c$ rather than the commonly used normalized efficiency $\eta=\Delta I_c/(I_{c+}+I_{c-})$, since $\eta$ can be artificially amplified when the denominator $I_{c+}+I_{c-}$ becomes small.
For $\zeta=0$, where the two arms are effectively independent, a sizable $\Delta I_c$ is confined to a narrow region with small $d\epsilon$ and $U$.
In addition, a weak diode signal appears along the tilted line $d\epsilon \approx 2U$ (thin dashed line), indicating a fragile diode window with strong parameter sensitivity.
When switching on the nonlocal channel, $\zeta = 1.0$ as shown in panel (b), the tilted feature remains, but a new and pronounced horizontal “diode band” emerges: large $\Delta I_c$ develops and persists over a broad interaction range.
This establishes a robust operating regime with strong absolute nonreciprocity.
Panel (c) plots $\Delta I_c$ as a function of the nonlocal coupling strength $\zeta$ at $U=5.0$.
The position of the horizontal diode band continuously shifts toward larger $d\epsilon$ as $\zeta$ increases.
The horizontal and tilted diode signals originate from the nonlocal and local $0$-$\pi$ phase transitions, respectively, as will be discussed in detail in the following sections.
It is also worth noting that the sign of $\Delta I_c$ in Fig.~\ref{fig: eta phase} can be reversed by swapping the gate polarity ($d\epsilon \to -d\epsilon$) or tuning the flux, enabling experimental control of the rectification direction in the SQUIDs.

\paragraph{Branch-selection mechanism --}

To reveal the microscopic origin of the diode effects in Fig.~\ref{fig: eta phase}, we consider a vertical cut at $U=5.0$ with the nonlocal coupling switched on ($\zeta = 1.0$).
Varying the detuning $d\epsilon$ along this line, we plot the critical-current asymmetry $\Delta I_c = I_{c+} - I_{c-}$ in Fig.~\ref{fig: mechanism}(a) (black solid curve in the leftmost panel).
As $d\epsilon$ increases, $\Delta I_c$ exhibits a pronounced positive peak (horizontal diode band), followed by a weak negative peak (tilted diode band).
For the first positive peak, the system experiences four different stages, marked as \circledtext{1} to \circledtext{4}.
The corresponding CPR is plotted in the middle panels: black solid line for the physical CPR, red and blue dashed lines for the current calculated by the GS energies constrained in the singlet and doublet sectors, showed in the inset via $I^{s/d} = 2 \partial E^{s/d}_{\rm GS} / \partial {\Delta\phi}$.

For $d\epsilon=0$, the singlet branch remains the GS throughout $\Delta\phi\in[-\pi,\pi]$.
Small detuning $d\epsilon$ in regime \circledtext{1} induces a doublet GS in narrow windows near $\Delta\phi \approx \pm \pi$.
In this regime, both $I_{c+}$ and $I_{c-}$ are still attained on the singlet branch, so the diode effect still originates from the detuning-induced skewing of the singlet CPR.
With $d\epsilon$ increasing, the system enters the second stage \circledtext{2}, where the singlet-stable intervals within $\Delta\phi\in[-\pi,\pi]$ shrink, and the phase extrema $\Delta\phi_\pm$ become pinned at the $0$-$\pi$ phase boundary.
This ``pinning'' strongly enhances the CPR asymmetry and significantly boosts the diode efficiency $\eta = \Delta I_c/(I_{c+} + I_{c-})$ (labeled atop the figure), serving as a precursor to the branch selected JDE.
Upon further increasing $d\epsilon$, a qualitative change occurs when $I_{c\pm}$ are located on the different branches: $I_{c+}$ is located on the singlet sector, whereas $I_{c-}$ is located on the doublet sector, as shown in \circledtext{3}.
We shaded this area in yellow color, where the maximal critical current asymmetry $\Delta I_c$ located, and we call this \emph{branch selected} JDE.
With further increase of $d\epsilon$, the singlet-interval quickly shrinks, and $\Delta I_c$ goes down rapidly.
Eventually, the system enters into the regime \circledtext{4} that both $I_{c\pm}$ are attained in the doublet sector, where diode effect thus comes from the skewing of the doublet CPR, and the $\Delta I_c$ rapidly diminishes in this regime.
The second negative peak of $\Delta I_c$ around $d\epsilon \approx 10.0$ experiences similar stages, governed by the branch-selection mechanism as masked by the yellow region, however, with the strength much weaker than the first peak.
The reason will be clear in the next section.

Panel (b) of Fig.~\ref{fig: mechanism} provides a global view of the GS sectors [see Eq.\eqref{eq: Hamiltonian sector}] that determine $I_{c+}$ and $I_{c-}$ in the ($d\epsilon$, $U$) plane.
The lightblue region indicates that the GS at both $I_{c+}$ and $I_{c-}$ belongs to the even-parity singlet sector ($\hat{H}^{\Pi =1}_{S_z=0}$), corresponding to the $0$-phase, whereas the grey region indicates that the GS at both $I_{c+}$ and $I_{c-}$ belongs to the odd-parity doublet sector ($\hat{H}^{\Pi = -1}_{|S_z|=1/2}$), i.e., the $\pi$-phase, which is doubly degenerate in the absence of a magnetic field. Most notably, the intermediate orange region marks the branch selected regime, where $I_{c-}$ is attained in the $\pi$-phase [$\hat{H}^{\Pi = -1}_{|S_z|=1/2}$], while $I_{c+}$ remains in the $0$-phase [$\hat{H}^{\Pi =1}_{S_z=0}$].
By comparing with the $\zeta = 1.0$ result in Fig.~\ref{fig: eta phase}, one can see strong JDE closely tracks this orange region, demonstrating that the robust diode response is tied to branch selection between singlet and doublet sectors.

\paragraph{Local vs nonlocal $0$-$\pi$ transition --}

\begin{figure}
\includegraphics[width=1.0\linewidth]{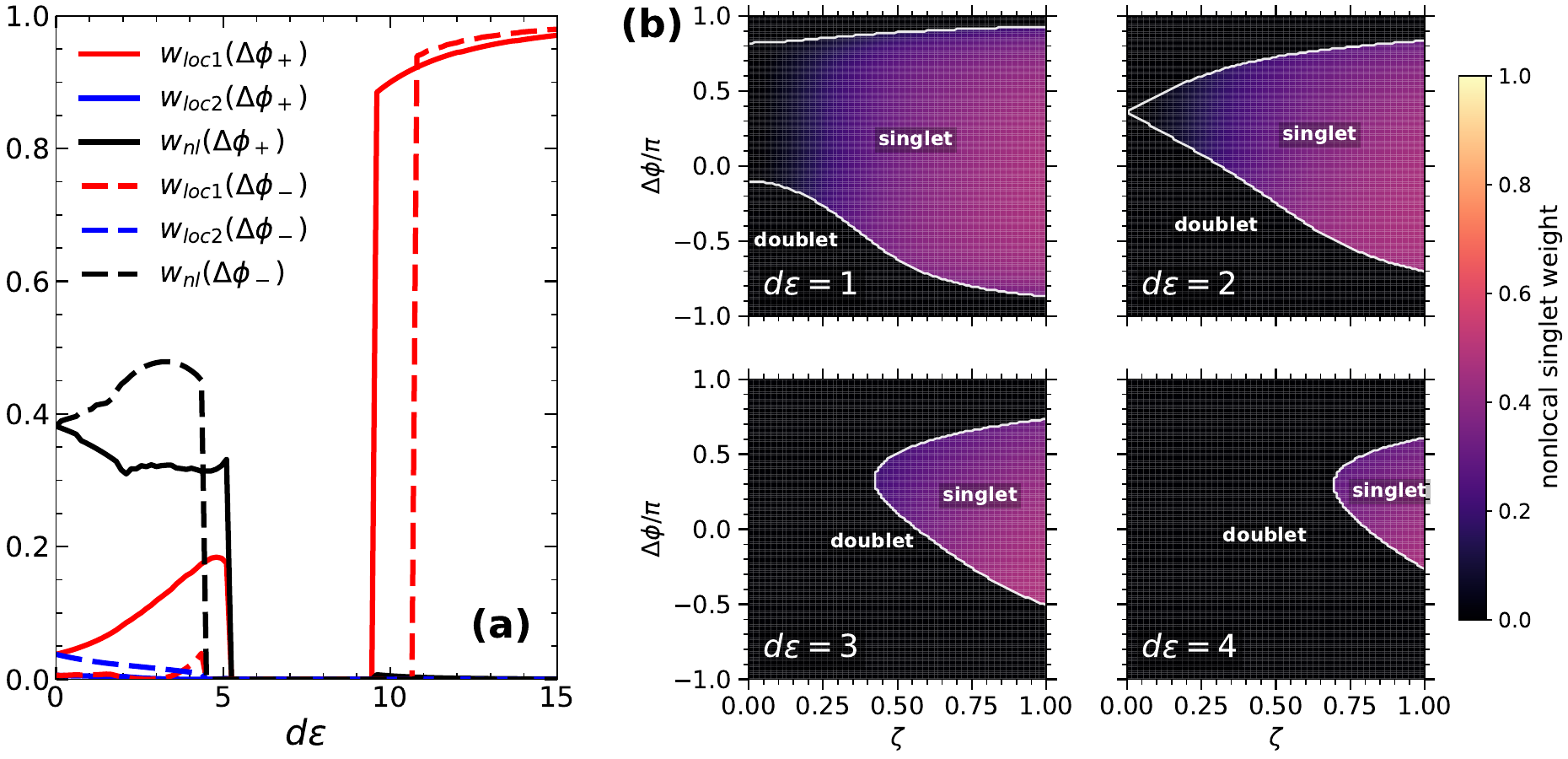}
\caption{(a) Projection of the GS ($\zeta = 1.0$) at $\Delta \phi_\pm$ onto the local $\ket{\uparrow\downarrow,0}$, $\ket{0,\uparrow\downarrow}$ and nonlocal $\frac{1}{\sqrt{2}}[\ket{\uparrow,\downarrow} - \ket{\downarrow,\uparrow}]$ singlet states.
(b) $0$-$\pi$ phase diagram in $(\Delta\phi, \zeta)$-plane with $U=5.0$ for various $d\epsilon$. The heatmap in the singlet phase indicates the weight of the nonlocal singlet state.}
\label{fig: nonlocal channel influence}
\end{figure}

To understand why the second peak in Fig.~\ref{fig: mechanism}(a) is weaker than the first one, although both originate from the same branch-selection mechanism, we analyze the character of the GS at the two critical phases $\Delta\phi_\pm$ where $I_{c\pm}$ are attained. 
Specifically, we project $\ket{\mathrm{GS}(\Delta\phi_\pm)}$ onto the local singlets $\ket{\rm{loc1}} = \ket{\uparrow\downarrow,0}$ and $\ket{\rm{loc2}} = \ket{0,\uparrow\downarrow}$ and the nonlocal singlet 
$\ket{\rm{nl}}=\left(\ket{\uparrow,\downarrow}-\ket{\downarrow,\uparrow}\right) / \sqrt{2}$, and evaluate the corresponding weights $w_\alpha(\Delta\phi_\pm)=\big|\langle \alpha \,|\, \mathrm{GS}(\Delta\phi_\pm)\rangle\big|^2$.

Fig.~\ref{fig: nonlocal channel influence}(a) shows these weights as a function of $d\epsilon$ at $U=5.0$, using the same parameters as in Fig.~\ref{fig: mechanism}(a).
For $d\epsilon\lesssim 5$, the singlet at $\Delta\phi_\pm$ carries a substantial nonlocal component (black), while the local component on dot $1$ (red) increases only moderately.
This identifies the pronounced horizontal ``diode band'' in Fig.~\ref{fig: eta phase} as being governed primarily by a \emph{nonlocal} $0$-$\pi$ transition.
In contrast, when the system re-enters the singlet region at large $d\epsilon$, the GS at $\Delta\phi_\pm$ becomes completely dominated by a \emph{local} paired configuration.
Accordingly, the weak tilted feature near $d\epsilon\simeq 2U$ is traced back to a \emph{local} $0$--$\pi$ transition in the single-dot-like limit.
This comparison shows that the nonlocal $0$-$\pi$ transition is much more effective in generating a strong diode response than the local one.
The position of this tilted line can be estimated analytically from a simple energy balance.
For $d\epsilon>0$, dot $1$ is active and the $0$-phase is well approximated by $\ket{\uparrow\downarrow,0}$ with energy $2\epsilon_1+U$, whereas the competing $\pi$-phase is a local-moment doublet $\ket{\uparrow,0}$ ($\ket{\downarrow,0}$) with energy $\epsilon_1$.
The crossing condition $E_{\ket{\uparrow\downarrow,0}} \approx E_{\ket{\uparrow,0}}$ yields $\epsilon_1\simeq -U$, i.e., $d\epsilon\simeq 2U$ under our symmetric parametrization $d\epsilon := -2\epsilon_1= 2\epsilon_2$.

Fig.~\ref{fig: nonlocal channel influence}(b) further maps out the $0$-$\pi$ phase boundary in the $(\Delta\phi,\zeta)$ plane for several representative $d\epsilon$, while the color scale within the singlet region indicates the weight of the nonlocal singlet component.
As $\zeta$ increases, the 0-phase region either expands ($d\epsilon=1,2$) or emerges ($d\epsilon=3,4$).
This behavior coincides with a rise in nonlocal weight (indicated by color), reflecting the enhanced admixture of the split-pair singlet $\ket{\text{nl}}$ into the even-parity sector, and the $0$-$\pi$ phase boundary is largely reshaped by the nonlocal pairing for large $\zeta$.
Because branch selection requires the CPR extrema $\Delta\phi_\pm$ to lie on opposite sides of the $0$-$\pi$ boundary, it is most readily realized when the $0$- and $\pi$-phase portions along $\Delta\phi$ become comparable, so that $\Delta\phi_\pm$ naturally straddle the crossing.
The reshaped $0$-$\pi$ boundaries in Fig.~\ref{fig: nonlocal channel influence}(b) directly account for the upward shift of the strong diode signal with increasing $\zeta$ in Fig.~\ref{fig: eta phase}(c), and ultimately give rise to the horizontal diode band in Fig.~\ref{fig: eta phase}(b).

\paragraph{Finite-gap robustness --}
Realistic superconducting gaps are typically on the meV scale, whereas electronic energy scales relevant for transport may extend up to the eV range.
At first sight, our superconducting atomic limit ($|\Delta|\to\infty$) might therefore appear remote from experimentally relevant conditions.
Nevertheless, the many-body diode mechanism discussed above remains applicable to realistic SQUIDs with a finite superconducting gap.
A simple way to demonstrate this is to employ the generalized atomic limit (GAL), in which the quasiparticle continuum is integrated out while retaining the leading finite-gap corrections \cite{Zonda2015,ifmmodeZelseZfionda2016,Kadlecova2019,ifmmodeZelseZfionda2023}.
Within this framework, the finite-gap problem can still be mapped onto an effective local Hamiltonian with the same Hamiltonian structure as in the atomic limit in Eq.~\eqref{eq: minimal model Hamiltonian}, but with renormalized parameters rescaled by $\nu$ and $\nu^2$, depending on whether they are multiplied by quadratic or quartic fermion operators in $\hat{H}^{\rm{eff}}$, where $\nu = 1 / \sqrt{1+\Gamma/\Delta}$ \cite{Zonda2015,ifmmodeZelseZfionda2023}.
This approach is known to reproduce phase boundary with high accuracy when benchmarked against numerically exact techniques such as the numerical renormalization group.
Importantly, once these renormalizations are accounted for, the qualitative features central to our conclusions, in particular, the branch selected many-body JDE and the nonlocal channel induced robustness are not artifacts of the $|\Delta|\to\infty$ approximation, but rather reflect generic interaction-driven physics of SQUIDs.

\paragraph{Conclusions --}
We have presented a many-body, branch-selection mechanism for a strong JDE in an interacting nano-SQUID.
The central concept relies on the non-analytic nature of the CPR across the $0$-$\pi$ quantum phase transition:
when the positive and negative critical currents are optimized on different many-body branches, a significant critical current asymmetry $\Delta I_c$ arises.
In our SQUID model, we find the maximum $\Delta I_c$ closely traces the branch-selection boundary in parameter space, providing strong support for this picture.
Moreover, a key ingredient for a robust diode operating regime is the nonlocal (pair-splitting) Cooper pair channel.
It reshapes the phase boundary by reducing the energy splitting between the doublet and the (nonlocal) singlet states, and thereby converts a narrow, parameter-sensitive diode hotspot into a robust diode band.

Future work on this SQUID platform should quantify finite-gap corrections and assess the roles of dissipation and nonequilibrium quasiparticles in branch selection.
More broadly, the mechanism presented here may extend to other systems exhibiting GS parity crossings, such as topological Josephson junctions.
Finally, our analysis suggests a practical design principle for many-body superconducting nonreciprocity: robust rectification is favored when the CPR extrema $\Delta\phi_\pm$ straddle the phase boundary, a condition most readily met in the vicinity of a quantum phase transition. 

\begin{acknowledgements}
Z. Z. and J. Z. contribute equally to this work.
Y. Z. acknowledges the support from the National Natural Science Foundation of China (NSFC) Grants No. 12274279 and the Shanghai Science and Technology Innovation Action Plan (Grant No. 24LZ1400800).
J.Y. acknowledges the NSFC (Grant No. 12504288) and the Science and Technology Commission of Shanghai Municipality (Grant No. 25ZR1402550).
\end{acknowledgements}

%

\clearpage
\onecolumngrid
\begin{center}
	\textbf{\large Many-body Josephson diode effect in superconducting quantum interferometers: Supplemental Material}
\end{center}

\setcounter{equation}{0}
\setcounter{figure}{0}
\setcounter{table}{0}
\setcounter{page}{1}
\setcounter{section}{0}
\makeatletter
\renewcommand{\theequation}{S\arabic{equation}}
\renewcommand{\thefigure}{S\arabic{figure}}
\renewcommand{\bibnumfmt}[1]{[S#1]}
\renewcommand{\citenumfont}[1]{S#1}

In this Supplemental Material, we formulate a microscopic transport theory for the superconducting quantum interferometer (SQUID) at the quantum mechanical level.
Starting from a double-dot device coupled in parallel to two superconducting leads, we derive the corresponding Green’s function formulation and show that, in the superconducting atomic limit, the model reduces to the effective Hamiltonian used in the main text [Eq.~\eqref{eq: minimal model Hamiltonian}].
This derivation also makes explicit the microscopic origin of both the local and nonlocal pairing channels relevant to the many-body Josephson diode effect.

\section{Microscopic transport theory for SQUIDs}

\begin{figure}[!b]
\includegraphics[width=0.6\linewidth]{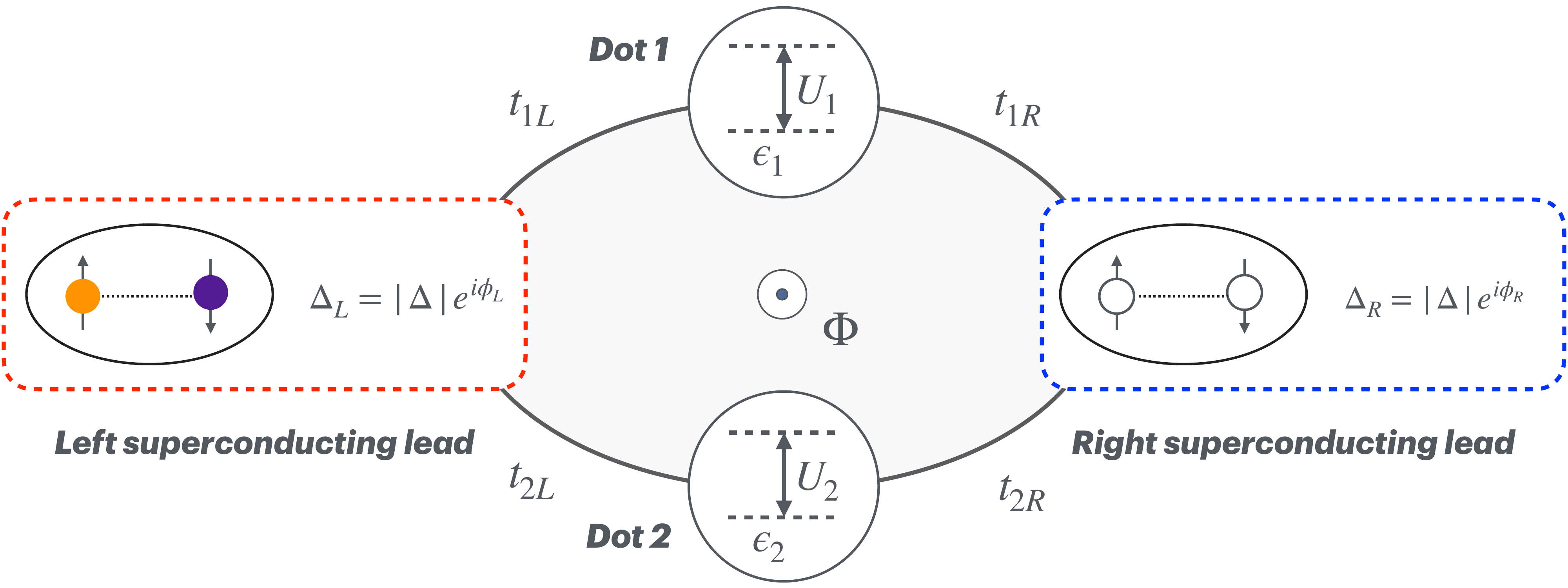}
\caption{Schematic of the SQUID model, consisting of two parallel quantum dots coupled between two superconducting leads. The leads are assumed to be identical s-wave superconductors, differing only in their macroscopic phases. An external magnetic field modulates the electron tunneling phases between the dots and the leads through the Peierls substitution.}
\label{fig: sm-model}
\end{figure}

\subsection{Model Hamiltonian}

Our SQUID model consists of two quantum dots connected in parallel between two superconducting leads, as illustrated in Fig.~\ref{fig: sm-model}.
The full Hamiltonian $\hat{H}$ naturally separates into three parts: (i) the two superconducting leads; (ii) the central double-dot device region; and (iii) the hybridization between the dots and the leads,
\begin{dmath}\label{eq: Hamiltonian squids}
\hat{H} = \sum_{i} \hat{H}_i^{dot} + \sum_{\alpha} \hat{H}_{\alpha}^{lead} + \sum_{\alpha i} \hat{H}^{hyb}_{\alpha i}~,
\end{dmath}
with
\begin{subequations}\label{eq: Hamiltonian squids detailed}
\begin{align}
\hat{H}_i^{dot} &=
\sum_\sigma (\epsilon_i - \sigma h_i) d^\dag_{i\sigma} d_{i\sigma}
+ U_i d^\dag_{i\uparrow} d_{i\uparrow} d^\dag_{i\downarrow} d_{i\downarrow}
~,\\
\hat{H}^{lead}_\alpha &=
\sum_{mn}
\sum_\sigma
T^\alpha_{mn} c^\dag_{\alpha m \sigma} c_{\alpha n \sigma}
+ \sum_m \left( \Delta_\alpha c^\dag_{\alpha m \uparrow} c^\dag_{\alpha m \downarrow} + h.c.\right)
~,\\
\hat{H}^{hyb}_{\alpha i} &=
-
\sum_{m \sigma} \left( t_{im\alpha} d_{i\sigma}^\dag c_{\alpha m \sigma} + h.c. \right) \qquad (t_{im\alpha} := t^{0}_{im\alpha} e^{i\theta_{i\alpha}})~.
\end{align}
\end{subequations}
Here, $i \in \{1,2\}$ labels the two quantum dots, and $\alpha \in \{L,R\}$ labels the left and the right leads.
The operators $d^{(\dagger)}_{i\sigma}$ and $c^{(\dagger)}_{\alpha m \sigma}$ annihilate (create) an electron with spin $\sigma$ on the $i$th quantum dot and on site $m$ of lead $\alpha$, respectively.
The parameters $\epsilon_i$, $h_i$, and $U_i$ denote the on-site energy, Zeeman splitting, and Coulomb charging energy of dot $i$, respectively.
We assume that the leads remain in local equilibrium, and $T^\alpha_{mn}$ denotes the tunneling matrix element within $\alpha$-lead from site $n$ to site $m$.
The local chemical potential of each lead is absorbed into the diagonal part of the hopping matrix, $T^\alpha_{mn} = T^0_{mn} + \delta_{mn}\epsilon_{\alpha}$, with $T^0_{mm}=0$ by convention.
The two superconducting leads are assumed to have the same gap amplitude and to differ only by their condensate phases, namely $\Delta_\alpha 
= |\Delta| e^{i\phi_\alpha}$.
Since only the superconducting phase difference, $\Delta\phi = \phi_L - \phi_R$, has physical significance, we adopt the gauge choice $\phi_L = +\Delta\phi/2$ and $\phi_R = -\Delta\phi/2$ throughout the following discussion.
The quantity $t_{im\alpha}$ denotes the tunneling amplitude from site $m$ of lead $\alpha$ to the dot $i$.
Its phase is modulated by the external magnetic field through the Peierls substitution, i.e.
\begin{equation}\label{eq: Peierls subtitution}
t_{\mathbf R_i \mathbf R_j} = t^0_{\mathbf R_i \mathbf R_j} e^{i\frac{q}{\hbar}\int_{\mathbf R_j}^{\mathbf R_i} \mathbf A(\mathbf r',t) \cdot d\mathbf r'} := t^0_{\mathbf R_i \mathbf R_j} e^{i\theta_{\mathbf R_i \mathbf R_j}}~,
\end{equation}
with electron charge $q=-e<0$.
$t^0$ denotes the amplitude in the absence of the magnetic field, which is assumed to be real.
In our SQUID geometry, we assume for simplicity that the tunneling amplitude from the dots to each site in the leads is uniform, such that $t_{im\alpha}$ reduces to $t_{i\alpha}$.
An electron encircling the loop acquires an Aharonov-Bohm phase $\frac{q\Phi}{\hbar} = - \frac{2\pi \Phi}{\Phi_e}$, where $\Phi_e = h/e$ is the normal-state flux quantum.
We choose a gauge in which the phases are distributed symmetrically along the loop, such that
\begin{equation}\label{eq: equally distributed gauge}
\theta_{1L} = -\theta_{\Phi}, \qquad
\theta_{2L} = +\theta_{\Phi}, \qquad
\theta_{1R} = +\theta_{\Phi}, \qquad
\theta_{2R} = -\theta_{\Phi},
\end{equation}
with $\theta_{\Phi} = \frac{q\Phi}{4\hbar} = -\frac{\pi\Phi}{4\Phi_0}$.
Here, $\Phi_0=\Phi_e/2=h/(2e)$ is the superconducting flux quantum.
Note that the factor of two arises because the superconducting order parameter describes a Cooper pair carrying charge $2e$.
Its phase winding around the loop is therefore twice the electronic Aharonov--Bohm phase.

Under the static mean-field description of the superconducting leads, particle number is no longer conserved because electrons can be converted into holes through pairing.
It is therefore convenient to work in Nambu space and introduce the following two-component spinors for both the dots and the leads:
\begin{equation}
\hat{\varphi}_i = 
\begin{pmatrix}
d_{i\uparrow} \\
d^\dag_{i\downarrow}
\end{pmatrix}~,
\qquad
\hat{\varphi}_i^\dag =
\begin{pmatrix}
d_{i\uparrow}^\dag & d_{i\downarrow}
\end{pmatrix}~,
\qquad
\hat{\psi}_{\alpha m} = 
\begin{pmatrix}
c_{\alpha m\uparrow} \\
c^{\dag}_{\alpha m\downarrow}
\end{pmatrix}~,
\qquad
\hat{\psi}_{\alpha m}^{\dag} =
\begin{pmatrix}
c_{\alpha m\uparrow}^{\dag} & c_{\alpha m\downarrow}
\end{pmatrix}~.
\end{equation}
The Hamiltonian thus can be rewritten in terms of the Nambu spinors as (up to a constant) \cite{Bruus2004_sm}
\begin{dmath}\label{eq: Hamiltonian in Nambu form}
\hat{H} = \sum_{i} \left( \underbrace{\hat{\varphi}_i^\dag \tilde{H}_i^0 \hat{\varphi}_i + U_i d^\dag_{i\uparrow} d_{i\uparrow} d^\dag_{i\downarrow} d_{i\downarrow}}_{:=\hat{H}_i^{dot}} \right)
+
\sum_{\alpha} \left( \underbrace{\sum_{mn} \hat{\psi}^\dag_{\alpha m} \tilde{H}^{\alpha}_{mn} \hat{\psi}_{\alpha n}}_{:=\hat{H}_\alpha^{lead}} \right)
+
\sum_{\alpha i} \left( \underbrace{-\sum_m \left( \hat{\varphi}_i^\dag \tilde{H}^{c\alpha}_{im} \hat{\psi}_{\alpha m} + \hat{\psi}_{\alpha m}^\dag \tilde{H}_{mi}^{\alpha c} \hat{\varphi}_i \right)}_{:=\hat{H}_{\alpha i}^{hyb}}\right)~,
\end{dmath}
where
\begin{equation}
\tilde{H}^0_i = 
\begin{pmatrix}
\epsilon_i - h_i & 0 \\
0 & - (\epsilon_i + h_i)
\end{pmatrix}~,\qquad
\tilde{H}^\alpha_{mn} = 
\begin{pmatrix}
T^\alpha_{mn} & \Delta_\alpha \delta_{mn} \\
\Delta_\alpha^* \delta_{mn} & -(T^\alpha_{mn})^*
\end{pmatrix}~,\qquad
\tilde{H}^{c \alpha}_{im} =
\begin{pmatrix}
t_{im\alpha} & \\
& -(t_{im\alpha})^*
\end{pmatrix}~.
\end{equation}
In the following, quantities carrying a tilde act in Nambu space and thus have an intrinsic $2 \times 2$ structure.
Note that $\tilde{H}^{\alpha c}_{mi} = (\tilde{H}_{im}^{c\alpha})^\dag$ in Eq.\eqref{eq: Hamiltonian in Nambu form} as required by Hermiticity.
Since the lead and device subspaces generally have different dimensions, the hybridization matrices $\tilde{H}^{\alpha c}$ and $\tilde{H}^{c\alpha}$ are in general rectangular rather than square.

\subsection{Nonequilibrium Green's function based transport theory}

We formulate the transport problem using Green’s functions.
Although the present work focuses on the equilibrium Josephson transport, we adopt the nonequilibrium Green’s function (NEGF) framework instead of the Matsubara formalism in order to maintain a more general theoretical formulation, as NEGF treats equilibrium and nonequilibrium situations on the same footing \cite{Stefanucci2013_sm,Haug2008_sm}.
The Nambu Green's function in the central scattering region is defined by
\begin{equation}\label{eq: Nambu Green's function}
\tilde{G}^{cc}_{ij}(z,z') = -i\avg{\mathcal{T}\hat{\varphi}_i(z) \otimes \hat{\varphi}^\dag_j(z')}
=-i
\begin{pmatrix}
\avg{\mathcal{T}d_{i\uparrow}(z)d^\dag_{j\uparrow}(z')} & \avg{\mathcal{T}d_{i\uparrow}(z)d_{j\downarrow}(z')} \\
\avg{\mathcal{T}d^\dag_{i\downarrow}(z)d^\dag_{j\uparrow}(z')} & \avg{\mathcal{T}d^\dag_{i\downarrow}(z)d_{j\downarrow}(z')}
\end{pmatrix}~,
\end{equation}
where $z$ and $z'$ are contour time variables that defined on the nonequilibrium Keldysh contour.
Indices $i$ and $j$ label the two dots.
The diagonal Nambu components in Eq.\eqref{eq: Nambu Green's function} describe normal electron and hole propagation, whereas the off-diagonal components describe anomalous propagation induced by superconducting pairing.

In the interacting problem, an exact evaluation of the full Green’s function is generally impossible because the
many-body Hilbert space grows exponentially.
A useful starting point is therefore the equation of motion for the central Green’s function (differential form):
\begin{dmath}\label{eq: differential form of the Dyson equation}
\sum_k \left( i\delta_{ik}\frac{\partial}{\partial z} - \tilde{H}_{ik}^0 \right) \tilde{G}^{cc}_{kj}(z,z') 
= \delta(z-z') \delta_{ij} + \sum_{\alpha =L,R} \left[ \tilde{\Sigma}^{\alpha} * \tilde{G}^{cc} \right]_{ij}(z,z')
+ \left[ \tilde{\Sigma}^{int} * \tilde{G}^{cc} \right]_{ij}(z,z')~.
\end{dmath}
Here,
\begin{dmath}\label{eq: SC lead self-energy}
\tilde{\Sigma}^{\alpha}_{ij}(z,z') = \sum_{mn} \tilde{H}^{c \alpha}_{im} \tilde{G}^{\alpha\alpha,0}_{mn}(z,z') \tilde{H}^{\alpha c}_{nj}~,
\end{dmath}
is the embedding self-energy from the $\alpha$-lead, which is obtained by integrating the lead degrees of freedom.
$\tilde{G}^{\alpha\alpha,0}$ in Eq.\eqref{eq: SC lead self-energy} is the Green's function of the isolated $\alpha$-lead, whose EoM takes a closed form
\begin{dmath}\label{eq: eom of the decoupled lead gf}
i\frac{\partial}{\partial z} \tilde{G}^{\alpha\alpha,0}_{mn}(z,z') = \delta(z-z')\delta_{mn} + \avg{\mathcal{T}\frac{\partial}{\partial z}\hat{\psi}_{\alpha m}(z) \hat{\psi}^\dag_{\alpha n}(z')}_0~.
\end{dmath}
$\tilde{\Sigma}^\alpha$ in Eq.\eqref{eq: SC lead self-energy}, in principle, can be exactly calculated by recursive Green's function method \cite{Godfrin1991_sm}.
However, when the detailed band structure of the leads does not qualitatively affect the low-energy physics, one may invoke the wide-band limit (WBL), in which an analytical expression for $\tilde{\Sigma}^\alpha$ becomes available.
The physical content of this approximation is that the leads act as broad featureless reservoirs on the energy scale relevant to the dots.
One therefore keeps the phase rigidity and pairing structure of the superconductors, but discards band-shape details that do not control the many-body interference mechanism of interest.

The Coulomb interaction induced many-body self-energy $\tilde{\Sigma}^{int}$ in Eq.\eqref{eq: differential form of the Dyson equation} is defined by
\begin{dmath}\label{eq: interaction self-energy}
[\tilde{\Sigma}^{int} * \tilde{G}^{cc}]_{ij}(z,z') = -iU_i
\begin{pmatrix}
+\avg{\mathcal{T} d_{i\uparrow}(z) d^\dag_{i\downarrow}(z^+) d_{i\downarrow}(z) d_{j\uparrow}^\dag(z')} & +\avg{\mathcal{T} d_{i\uparrow}(z) d^\dag_{i\downarrow}(z^+) d_{i\downarrow}(z) d_{j\downarrow}(z')} \\
-\avg{\mathcal{T} d^\dag_{i\uparrow}(z^+) d_{i\uparrow}(z) d^\dag_{i\downarrow}(z) d_{j\uparrow}^\dag(z')} & -\avg{\mathcal{T} d^\dag_{i\uparrow}(z^+) d_{i\uparrow}(z) d^\dag_{i\downarrow}(z) d_{j\downarrow}(z')}
\end{pmatrix}~.
\end{dmath}
In principle, $\tilde{\Sigma}^{int}$ can be treated by many-body perturbation theory \cite{Zonda2015_sm,Schwab1999_sm,Janiifmmodecheckselsevsfi2021_sm}, quantum Monte Carlo \cite{Siano2004_sm,Luitz2012_sm,Pokorny2021_sm}, numerical renormalization group \cite{Yoshioka2000_sm,Karrasch2008_sm,ifmmodeZelseZfiitko2010_sm}, and related methods.
In the present work, however, we do not attempt to compute this interaction self-energy directly.
Instead, we focus on the superconducting atomic limit $|\Delta| \to \infty$ (see next section), where the lead-induced self-energy becomes static and the problem reduces to an effective finite-dimensional Hamiltonian that can be diagonalized exactly.
This limit is especially valuable because it retains the essential competition between interaction, Zeeman splitting, magnetic flux, and proximity-induced pairing, while removing the dynamical complexity associated with quasiparticle continua in the leads.

Once the Green’s function, as well as the lead and interaction self-energies, have been obtained, the Josephson current flowing through lead $\alpha$ can be defined as
\begin{dmath}\label{eq: Josephson current definition}
I^{\alpha}(t) = -e \sum_{m\sigma} \frac{\partial}{\partial t} \avg{\hat{n}_{\alpha m \sigma}(t)}~,
\end{dmath}
where the minus sign arises from the negative electron charge.
In equilibrium, the current is conserved and independent of time. We thus define the stationary current as $I=-I_L=I_R$ with positive current taken to flow from the left lead to the right lead.
One can employ the Heisenberg equation of motion to simplify Eq.~\eqref{eq: Josephson current definition}, and notice that only the anti-commutation of $[\hat{H}^{hyb}_{\alpha i}, \hat{N}_\alpha]$ contributes to Eq.~\eqref{eq: Josephson current definition} \footnote{Mathematically, $[\hat H^{\mathrm{lead}}_\alpha,\hat{N}_\alpha]\neq 0$ because the BCS mean-field Hamiltonian in the superconducting lead does not conserve the electron number. However, this term describes only the internal pair-conversion dynamics within lead $\alpha$, rather than particle transfer across the lead-central-region interface. The transport current is defined by the interfacial charge transfer and is therefore determined solely by the hybridization term $\hat{H}^{\mathrm{hyb}}_\alpha$.}.
After transforming to frequency space, the current can finally be written in the Meir-Wingreen form:
\begin{dmath}\label{eq: charge current formula}
I = \frac{e}{\hbar} \sum_{ij} \Tr_N \int_{-\infty}^{+\infty} \frac{d\omega}{2\pi} \left[ \tilde{\tau}_3 \left( \tilde{\Sigma}^{L,r}_{ij}(\omega) \tilde{G}^{cc,<}_{ji}(\omega) + \tilde{\Sigma}^{L,<}_{ij}(\omega) \tilde{G}^{cc,a}_{ji}(\omega) \right) \right] + h.c.~,
\end{dmath}
where $\tilde{\tau}_3 = diag[+1,-1]$, the third Pauli matrix, and $\Tr_N$ represents the trace over the Nambu space.
It is important to note that the above current formula is equivalent to Eq.~\eqref{eq: Josephson current} in the main text, namely, $I = \frac{2e}{\hbar} \frac{\partial \Omega}{\partial \Delta\phi}$, provided that the grand-canonical potential of the system can be evaluated exactly.
A simple illustration of this equivalence is given by the superconducting atomic limit, as discussed in the next section.

So far, we've transformed the semi-infinite long leads into the lead self-energies, and the transport problem is therefore reduced entirely to the finite central region.
In what follows, unless explicitly stated otherwise, we will omit the 'cc' superscript without risk of ambiguity.

\subsection{Wide band limit approximation for s-wave superconducting leads}

If the detailed band structure of the superconducting leads is not important for the qualitative physics, one can apply the WBL approximation and derive an analytical expression for the lead self-energy.
Recall that the leads are assumed in local equilibrium, so that only the retarded component needs to be computed as the `lesser' component can be obtained by using the fluctuation-dissipation theorem  \cite{Haug2008_sm}, namely $\tilde{\Sigma}^{\alpha,<}(\omega;\kvec) = i f(\omega) \tilde{\Gamma}^{\alpha}(\omega;\kvec)$ where $\tilde{\Gamma}^{\alpha}(\omega;\kvec) = i[\tilde{\Sigma}^{\alpha,r}(\omega;\kvec) - \tilde{\Sigma}^{\alpha,a}(\omega;\kvec)]$.
By applying Langreth's rules on Eq.\eqref{eq: SC lead self-energy}, we have ($m,n$ are restricted in the unit cell after the Fourier transform)
\begin{dmath}\label{eq: SC lead self-energy -- retarded}
\tilde{\Sigma}^{\alpha,r}_{ij}(\omega) = \frac{1}{N_\kvec} \sum_{\kvec} \tilde{H}^{c \alpha}_{im}(\kvec) \tilde{G}^{\alpha\alpha,0,r}_{mn}(\omega;\kvec) \tilde{H}^{\alpha c}_{nj}(\kvec)~.
\end{dmath}
The Fourier transformed decoupled lead Green's function in Eq.\eqref{eq: eom of the decoupled lead gf} reads
\begin{equation}
\tilde{G}^{\alpha\alpha,0,r}(\omega;\kvec) = 
\begin{pmatrix}
\omega^+ - \epsilon(\kvec)  & \Delta_\alpha  \\
\Delta_\alpha^* & \omega^+ + \epsilon(-\kvec)
\end{pmatrix}^{-1}
=
\frac{1}{(\omega^+)^2 - [\epsilon(\kvec)]^2 - |\Delta|^2}
\left(
\begin{matrix}
\omega^+ + \epsilon(\kvec)& -\Delta_\alpha & \\
-\Delta_\alpha^* & \omega^+ - \epsilon(\kvec)
\end{matrix}
\right)~,
\end{equation}
where we assume $\epsilon(\kvec) = \epsilon(-\kvec)$ in the second equality as the leads having inversion symmetry.
As a result,
\begin{equation}\label{eq: SC lead self-energy -- retarded 2}
\tilde{\Sigma}^{\alpha,r}_{ij}(\omega) = 
\int_{-\infty}^\infty dE
\frac{\rho(E)}{(\omega^+)^2 - E^2 - | \Delta |^2}
\begin{pmatrix}
{t}_{i\alpha} ({t}_{j\alpha})^* (\omega^+ + \cancel{E}) & \Delta_\alpha {t}_{i\alpha} {t}_{j\alpha} \\
[\Delta_\alpha {t}_{i\alpha} {t}_{j\alpha}]^* & ({t}_{i\alpha})^* {t}_{j\alpha} (\omega^+ - \cancel{E})
\end{pmatrix}~,
\end{equation}
where $\rho(E) = \frac{1}{N_\kvec} \sum_{\kvec} \delta(E - \epsilon(\kvec))$ is the normal state density of states (DoS) of the lead.
The terms proportional to $E$ in the diagonal entries vanish after integration because the integrand is odd in $E$.
In WBL, we assume the DoS is flat that ranges in $\omega \in [-D,+D]$ with value $\bar{\rho} = 1/2D$.
From Eq.\eqref{eq: Peierls subtitution}, we write the dot-lead tunneling amplitude as $t_{i\alpha}=t^0 e^{i\theta_{i\alpha}}$, where $t^0$ is the tunneling between the leads and the dots in the absence of the magnetic field.
We further define the hybridization strength $\Gamma = \pi | t^0 |^2 \bar{\rho}$.
As a result, the lead self-energy becomes
\begin{equation}\label{eq: lead self-energy -- wide band limit}
\tilde{\Sigma}^{\alpha,r}_{ij}(\omega)
=
s^r(\omega)
\begin{pmatrix}
\omega e^{i(\theta_{i\alpha} - \theta_{j\alpha})} & \Delta_\alpha e^{i(\theta_{i\alpha}+\theta_{j\alpha})} \\
[\Delta_\alpha e^{i(\theta_{i\alpha}+\theta_{j\alpha})}]^* & \omega e^{-i(\theta_{i\alpha} - \theta_{j\alpha})}
\end{pmatrix}
\quad
\text{with}
\quad
s^r(\omega) = \frac{\Gamma}{\pi} \int_{-D}^{+D} dE \frac{1}{(\omega^+)^2 - E^2 - |\Delta|^2}~.
\end{equation}
$s^r(\omega)$ can be evaluated directly, giving
\begin{equation}
s^r(\omega)
=
\begin{cases}
-\frac{2\Gamma}{\pi \sqrt{|\Delta|^2 - \omega^2}} \arctan \frac{D}{\sqrt{|\Delta|^2 - \omega^2}}\qquad&(\omega^2 < |\Delta|^2)~,\\
\frac{\Gamma}{\pi \sqrt{\omega^2 - |\Delta|^2}} \ln \left| \frac{D+\sqrt{\omega^2 - |\Delta|^2}}{-D + \sqrt{\omega^2 - |\Delta|^2}} \right| - \frac{i\Gamma sgn(\omega)}{\sqrt{\omega^2 - |\Delta|^2}} \theta(D - \sqrt{\omega^2 - |\Delta|^2})\qquad&(\omega^2 > |\Delta|^2)~.
\end{cases}
\end{equation}
In the limit $D \to \infty$, $s^r(\omega)$ is further simplified to \cite{Janiifmmodecheckselsevsfi2021_sm}
\begin{dmath}
\lim_{D \rightarrow +\infty} s^r(\omega) = 
\begin{cases}
-\frac{\Gamma}{\sqrt{|\Delta|^2 - \omega^2}} & (\omega^2 < |\Delta|^2)~,\\
-\frac{i\Gamma sgn(\omega)}{\sqrt{\omega^2 - |\Delta|^2}} & (\omega^2 > |\Delta|^2)~.
\end{cases}
\end{dmath}

\subsection{Local and nonlocal transport channel in SQUIDs}

The lead self-energy Eq.\eqref{eq: lead self-energy -- wide band limit} takes the following explicit $4 \times 4$ form if one first iterates the Nambu index and then the site index:
\begin{dmath}\label{eq: lead self-energy for squids}
\tilde{\Sigma}^{\alpha,r}(\omega) = s^r(\omega)
\left(
\begin{array}{c|c}
\begin{matrix}
\omega & |\Delta| e^{i(2\theta_{1\alpha}+\phi_\alpha)}\\
|\Delta| e^{-i(2\theta_{1\alpha}+\phi_\alpha)} & \omega
\end{matrix}
&
\begin{matrix}
\omega e^{i(\theta_{1\alpha}-\theta_{2\alpha})} & |\Delta| e^{i\phi_\alpha}\\
|\Delta| e^{-i\phi_\alpha} & \omega e^{-i(\theta_{1\alpha}-\theta_{2\alpha})}
\end{matrix}
\\[0.6em]
\hline
\\[-0.8em]
\begin{matrix}
\omega e^{-i(\theta_{1\alpha}-\theta_{2\alpha})} & |\Delta| e^{i\phi_\alpha}\\
|\Delta| e^{-i\phi_\alpha} & \omega e^{i(\theta_{1\alpha}-\theta_{2\alpha})}
\end{matrix}
&
\begin{matrix}
\omega & |\Delta| e^{i(2\theta_{2\alpha}+\phi_\alpha)}\\
|\Delta| e^{-i(2\theta_{2\alpha}+\phi_\alpha)} & \omega
\end{matrix}
\end{array}
\right)~.
\end{dmath}
The block diagonal terms describe local proximity pairing on each dot, while the block off-diagonal terms encode nonlocal pairing between the two dots.
If the interdot coherence is neglected, namely by setting the block off-diagonal part to zero, Eq.~\eqref{eq: charge current formula} reduces to
\begin{dmath}\label{eq: charge current formula -- decoherence}
I = \frac{e}{\hbar} \sum_{i} \Tr_N
\int_{-\infty}^{+\infty}
\frac{d\omega}{2\pi}
\left[
\tilde{\tau}_{3} \left(\tilde{\Sigma}_{ii}^{L,r}(\omega) \tilde{G}_{ii}^{cc,<}(\omega) + \tilde{\Sigma}_{ii}^{L,<}(\omega) \tilde{G}_{ii}^{cc,a}(\omega) \right)
\right] + h.c.~.
\end{dmath}
The total current through the SQUID then reduces to the sum of two independent Josephson currents, $I = I_{i=1} + I_{i=2}$.
From Eq.~\eqref{eq: lead self-energy for squids}, it follows that junctions 1 and 2 are associated with the effective phase differences $4\theta_{1\alpha}+2\phi_\alpha$ and $4\theta_{2\alpha}+2\phi_\alpha$, respectively. In the symmetric gauge of Eq.~\eqref{eq: equally distributed gauge}, the current in the absence of coherence between the two arms becomes \cite{Wang2011_sm}
\begin{dmath}\label{eq: charge current formula -- decoherence 2}
I(\Delta\phi,\Phi)
= I_{i=1} \left(\Delta\phi + \pi\frac{\Phi}{\Phi_0}\right)
+ I_{i=2} \left(\Delta\phi - \pi\frac{\Phi}{\Phi_0}\right)~\qquad\text{(without inter-dot coherence)}.
\end{dmath}
This explicitly shows that the phase differences seen by the two arms differ by $2\pi\Phi/\Phi_0$, which has been used in recent studies of SQUIDs \cite{Souto2022_sm,Qi2025_sm}.
By contrast, our derivation reveals that retaining the inter-arm coherence naturally gives rise to a nonlocal transport channel, which is essential for the robust Josephson diode effect highlighted in the main text.

\section{Superconducting atomic limit}\label{subsec:block-summary}

The above analysis provides a general transport formulation for the SQUID device.
The main difficulty in solving the problem lies in the interaction self-energy introduced in Eq.\eqref{eq: interaction self-energy}, which is, in general, frequency dependent and must be determined self-consistently within a many-body treatment. 
To expose the underlying physics more transparently, here we focus on the superconducting atomic limit $|\Delta|\to\infty$ \cite{Meng2009_sm}. In this limit, the quasiparticle continuum of the leads is projected out, while the essential competition among Coulomb interaction, Zeeman splitting, magnetic flux, and proximity-induced pairing is fully retained.
The total lead self-energy $\tilde{\Sigma}^{tot,r}(\omega) = \tilde{\Sigma}^{L,r}(\omega) + \tilde{\Sigma}^{R,r}(\omega)$ in the symmetric gauge thus becomes
\begin{dmath}\label{eq: total lead self-energy for squids}
\tilde{\Sigma}^{tot,r}(\omega) = 2 s^r(\omega)
\begin{pmatrix}
\omega & |\Delta|\cos\left( \frac{\Delta\phi}{2} + \frac{\pi\Phi}{2\Phi_0} \right) & \omega\cos\left(\frac{\pi\Phi}{2\Phi_0}\right) & |\Delta|\cos\left(\frac{\Delta\phi}{2}\right)\\
|\Delta|\cos\left(\frac{\Delta\phi}{2}+\frac{\pi\Phi}{2\Phi_0}\right) & \omega & |\Delta|\cos\left(\frac{\Delta\phi}{2}\right) & \omega\cos\left(\frac{\pi\Phi}{2\Phi_0}\right)\\
\omega\cos\left(\frac{\pi\Phi}{2\Phi_0}\right) & |\Delta|\cos\left(\frac{\Delta\phi}{2}\right) & \omega & |\Delta|\cos\left(\frac{\Delta\phi}{2}-\frac{\pi\Phi}{2\Phi_0}\right)\\
|\Delta|\cos\left(\frac{\Delta\phi}{2}\right) & \omega\cos\left(\frac{\pi\Phi}{2\Phi_0}\right) & |\Delta|\cos\left(\frac{\Delta\phi}{2}-\frac{\pi\Phi}{2\Phi_0}\right) & \omega
\end{pmatrix}
\end{dmath}

A major simplification occurs in the limit $|\Delta|\to\infty$.
In this limit, the frequency dependence of the lead self-energy disappears: the diagonal (normal) components vanish as $\omega/|\Delta|$, whereas the off-diagonal (anomalous) components approach finite constants. 
As a result, the superconducting leads generate only static instantaneous pairing terms in the dot subspace, which reduces to
\begin{dmath}\label{eq: total lead self-energy for squids -- infinite gap limit}
\tilde{\Sigma}^{tot,r}(\omega) = -2
\begin{pmatrix}
0 & \Gamma\cos\left( \frac{\Delta\phi}{2} + \frac{\pi\Phi}{2\Phi_0} \right) & 0 & \Gamma\cos\left(\frac{\Delta\phi}{2}\right)\\
\Gamma\cos\left(\frac{\Delta\phi}{2}+ \frac{\pi\Phi}{2\Phi_0}\right) & 0 & \Gamma\cos\left(\frac{\Delta\phi}{2}\right) & 0\\
0 & \Gamma\cos\left(\frac{\Delta\phi}{2}\right) & 0 & \Gamma\cos\left(\frac{\Delta\phi}{2} - \frac{\pi\Phi}{2\Phi_0}\right)\\
\Gamma\cos\left(\frac{\Delta\phi}{2}\right) & 0 & \Gamma\cos\left(\frac{\Delta\phi}{2} - \frac{\pi\Phi}{2\Phi_0}\right) & 0
\end{pmatrix}~.
\end{dmath}
Hence, in the $|\Delta| \to \infty$ limit, the static lead self-energy can be represented by an effective instantaneous pairing term in the dot subspace, which yields the effective Hamiltonian:
\begin{dmath}
\hat{H}^{\rm{eff}} =
\sum_{i} \left[ \sum_\sigma (\epsilon_i - \sigma h_i) d^\dag_{i\sigma} d_{i\sigma}
+ U_i d^\dag_{i\uparrow} d_{i\uparrow} d^\dag_{i\downarrow} d_{i\downarrow}
 \right]
-2\Gamma \cos\left(\frac{\Delta\phi}{2}+\frac{\pi\Phi}{2\Phi_0}\right) \left(d_{1\uparrow}^\dag d_{1\downarrow}^\dag + h.c.\right)
-2\Gamma \cos\left(\frac{\Delta\phi}{2}-\frac{\pi\Phi}{2\Phi_0}\right) \left(d_{2\uparrow}^\dag d_{2\downarrow}^\dag + h.c.\right)
-2\Gamma 
\cos(\frac{\Delta\phi}{2}) \left(d_{1\uparrow}^\dag d_{2\downarrow}^\dag + d_{2\uparrow}^\dag d_{1\downarrow}^\dag + h.c. \right)~,
\end{dmath}
known as superconducting atomic limit.
This Hamiltonian is exactly what we introduced in Eq.~\eqref{eq: minimal model Hamiltonian} in the main text with $\zeta = 1$.
In general, $\zeta \in [0,1]$ is a phenomenological parameter that accounts for the suppression of the nonlocal channel when the arm separation $d$ exceeds the superconducting coherence length $\xi$; in that regime the nonlocal pairing amplitude decays as $\zeta \sim e^{-d/\xi}$, and the microscopic limit $\zeta = 1$ is recovered for $d \ll \xi$ \cite{Wang2011_sm}.

Because $\hat{H}^{\rm{eff}}$ contains only density terms and spin-singlet pairing terms, both fermion parity $\hat{\Pi} = (-1)^{\hat{N}}$ and spin-z projection $\hat{S}_z$ are conserved.
The Hamiltonian can therefore be decomposed into the following sectors:
\begin{dmath}
H =
\mathrm{diag} 
\Big(
H^{\Pi=1}_{S_z=0},
H^{\Pi=1}_{S_z=+1},
H^{\Pi=1}_{S_z=-1},
H^{\Pi=-1}_{S_z=+1/2},
H^{\Pi=-1}_{S_z=-1/2}
\Big)~,
\end{dmath}
with the explicit block forms given by:
\begin{equation}\label{eq:H_reordered_16x16_tabular_labeled}
\setlength{\tabcolsep}{2.2pt}
\renewcommand{\arraystretch}{1.05}
\left(
\begin{tabular}{c|*{6}{c}|c|c|*{4}{c}|*{4}{c}}
%
 & $\ket{0,0}$ & $\ket{\uparrow\downarrow,0}$ & $\ket{0,\uparrow\downarrow}$ & $\ket{\uparrow,\downarrow}$ & $\ket{\downarrow,\uparrow}$ & $\ket{\uparrow\downarrow,\uparrow\downarrow}$
 & $\ket{\uparrow,\uparrow}$ & $\ket{\downarrow,\downarrow}$
 & $\ket{\uparrow,0}$ & $\ket{0,\uparrow}$ & $\ket{\uparrow\downarrow,\uparrow}$ & $\ket{\uparrow,\uparrow\downarrow}$
 & $\ket{\downarrow,0}$ & $\ket{0,\downarrow}$ & $\ket{\uparrow\downarrow,\downarrow}$ & $\ket{\downarrow,\uparrow\downarrow}$\\
\hline
$\ket{0,0}$ &
$0$ & $\alpha_1$ & $\alpha_2$ & $\beta$ & $-\beta$ & $0$ &
$0$ & $0$ &
$0$ & $0$ & $0$ & $0$ &
$0$ & $0$ & $0$ & $0$\\

$\ket{\uparrow\downarrow,0}$ &
$\alpha_1$ & $p$ & $0$ & $0$ & $0$ & $\alpha_2$ &
$0$ & $0$ &
$0$ & $0$ & $0$ & $0$ &
$0$ & $0$ & $0$ & $0$\\

$\ket{0,\uparrow\downarrow}$ &
$\alpha_2$ & $0$ & $q$ & $0$ & $0$ & $\alpha_1$ &
$0$ & $0$ &
$0$ & $0$ & $0$ & $0$ &
$0$ & $0$ & $0$ & $0$\\

$\ket{\uparrow,\downarrow}$ &
$\beta$ & $0$ & $0$ & $s$ & $0$ & $-\beta$ &
$0$ & $0$ &
$0$ & $0$ & $0$ & $0$ &
$0$ & $0$ & $0$ & $0$\\

$\ket{\downarrow,\uparrow}$ &
$-\beta$ & $0$ & $0$ & $0$ & $t$ & $\beta$ &
$0$ & $0$ &
$0$ & $0$ & $0$ & $0$ &
$0$ & $0$ & $0$ & $0$\\

$\ket{\uparrow\downarrow,\uparrow\downarrow}$ &
$0$ & $\alpha_2$ & $\alpha_1$ & $-\beta$ & $\beta$ & $z$ &
$0$ & $0$ &
$0$ & $0$ & $0$ & $0$ &
$0$ & $0$ & $0$ & $0$\\
\hline
$\ket{\uparrow,\uparrow}$ &
$0$ & $0$ & $0$ & $0$ & $0$ & $0$ &
$r$ & $0$ &
$0$ & $0$ & $0$ & $0$ &
$0$ & $0$ & $0$ & $0$\\
\hline
$\ket{\downarrow,\downarrow}$ &
$0$ & $0$ & $0$ & $0$ & $0$ & $0$ &
$0$ & $u$ &
$0$ & $0$ & $0$ & $0$ &
$0$ & $0$ & $0$ & $0$\\
\hline
$\ket{\uparrow,0}$ &
$0$ & $0$ & $0$ & $0$ & $0$ & $0$ &
$0$ & $0$ &
$a$ & $0$ & $-\beta$ & $\alpha_2$ &
$0$ & $0$ & $0$ & $0$\\

$\ket{0,\uparrow}$ &
$0$ & $0$ & $0$ & $0$ & $0$ & $0$ &
$0$ & $0$ &
$0$ & $c$ & $\alpha_1$ & $-\beta$ &
$0$ & $0$ & $0$ & $0$\\

$\ket{\uparrow\downarrow,\uparrow}$ &
$0$ & $0$ & $0$ & $0$ & $0$ & $0$ &
$0$ & $0$ &
$-\beta$ & $\alpha_1$ & $x$ & $0$ &
$0$ & $0$ & $0$ & $0$\\

$\ket{\uparrow,\uparrow\downarrow}$ &
$0$ & $0$ & $0$ & $0$ & $0$ & $0$ &
$0$ & $0$ &
$\alpha_2$ & $-\beta$ & $0$ & $v$ &
$0$ & $0$ & $0$ & $0$\\
\hline
$\ket{\downarrow,0}$ &
$0$ & $0$ & $0$ & $0$ & $0$ & $0$ &
$0$ & $0$ &
$0$ & $0$ & $0$ & $0$ &
$b$ & $0$ & $-\beta$ & $\alpha_2$\\

$\ket{0,\downarrow}$ &
$0$ & $0$ & $0$ & $0$ & $0$ & $0$ &
$0$ & $0$ &
$0$ & $0$ & $0$ & $0$ &
$0$ & $d$ & $\alpha_1$ & $-\beta$\\

$\ket{\uparrow\downarrow,\downarrow}$ &
$0$ & $0$ & $0$ & $0$ & $0$ & $0$ &
$0$ & $0$ &
$0$ & $0$ & $0$ & $0$ &
$-\beta$ & $\alpha_1$ & $y$ & $0$\\

$\ket{\downarrow,\uparrow\downarrow}$ &
$0$ & $0$ & $0$ & $0$ & $0$ & $0$ &
$0$ & $0$ &
$0$ & $0$ & $0$ & $0$ &
$\alpha_2$ & $-\beta$ & $0$ & $w$
\end{tabular}
\right)~,
\end{equation}
Here, each matrix element are given by
\begin{equation}
\begin{aligned}
a &= \epsilon_1 - h_1, &\quad
b &= \epsilon_1 + h_1, &\quad
c &= \epsilon_2 - h_2, &\quad
d &= \epsilon_2 + h_2, \\[4pt]
p &= 2\epsilon_1 + U_1, &\quad
q &= 2\epsilon_2 + U_2, &\quad
r &= a + c, &\quad
s &= a + d, \\[4pt]
t &= b + c, &\quad
u &= b + d, &\quad
v &= a + q, &\quad
w &= b + q, \\[4pt]
x &= p + c, &\quad
y &= p + d, &\quad
z &= p + q, \\[4pt]
\alpha_1 &= -2\Gamma \cos \left(\frac{\Delta\phi}{2}+\frac{\pi\Phi}{2\Phi_0}\right), &\quad
\alpha_2 &= -2\Gamma \cos \left(\frac{\Delta\phi}{2}-\frac{\pi\Phi}{2\Phi_0}\right), &\quad
\beta &= -2\Gamma \cos \left(\frac{\Delta\phi}{2}\right).
\end{aligned}
\end{equation}

One can, in principle, exactly diagonalize the Hamiltonian and obtain all the physical observables, according to $O = \frac{1}{Z}\sum_n e^{-\beta E_n} \avg{n | \hat{O} | n}$, with the partition function $Z = \sum_n e^{-\beta E_n}$.
Most importantly, in the superconducting atomic limit $|\Delta|\to\infty$, the current formula in Eq.~(S13) reduces to
\begin{equation}\label{eq: current in infinite gap limit}
I =
-\frac{2ie\Gamma}{\hbar}
\sum_{ij}
\left(
e^{i\phi_L} e^{i(\theta_{iL}+\theta_{jL})}
\langle d^\dagger_{i\uparrow} d^\dagger_{j\downarrow}\rangle
- h.c.
\right)~.
\end{equation}
Here we used that, in this limit, the lead self-energy, Eq.\eqref{eq: lead self-energy -- wide band limit}, becomes static and purely anomalous,
\begin{equation}
\tilde{\Sigma}^{\alpha,r}_{ij}
=
\tilde{\Sigma}^{\alpha,a}_{ij}
=
-\Gamma
\begin{pmatrix}
0 & e^{i\phi_\alpha}e^{i(\theta_{i\alpha}+\theta_{j\alpha})}\\
e^{-i\phi_\alpha}e^{-i(\theta_{i\alpha}+\theta_{j\alpha})} & 0
\end{pmatrix},
\end{equation}
so that $\tilde{\Sigma}^{\alpha,<}=0$.
One can then verify directly that Eq.\eqref{eq: current in infinite gap limit} is equivalent to Eq.\eqref{eq: Josephson current} in the main text:
\begin{equation}
I =
\frac{2e}{\hbar}
\left\langle
\frac{\partial \hat H_{\rm eff}}{\partial \phi_L}
\right\rangle
= \frac{2e}{\hbar}\frac{\partial F}{\partial \phi_L}~,
\end{equation}
with the second equality follows from the Hellmann-Feynman theorem.

\end{document}